\documentclass[12pt]{iopart}
\usepackage{epsfig}
\usepackage{amssymb}
\usepackage{mathrsfs}
\def\shiftleft#1{#1\llap{#1\hskip 0.04em}}
\def\shiftdown#1{#1\llap{\lower.04ex\hbox{#1}}}
\def\thick#1{\shiftdown{\shiftleft{#1}}}
\def\b#1{\thick{\hbox{$#1$}}}
\begin{document}

\title[Scalar three-body force and properties of the three-nucleon system]
{The properties of the three-nucleon system with the dressed-bag model
 for nn interaction. I: New scalar three-body force
}

\author{V. I. Kukulin, V. N. Pomerantsev}
 \address {Institute of Nuclear Physics, Moscow State University,
119899 Moscow, Russia}
\author{M. Kaskulov }
\address {Physikalisches Institut, Universit\"at T\"ubingen,\\ 
Auf der Morgenstelle 14, D-72076 T\"{u}bingen, Germany}
\author{and Amand Faessler }
\address {Institut f\"ur Theoretische Physik, Universit\"at T\"ubingen,\\ 
Auf der Morgenstelle 14, D-72076 T\"{u}bingen, Germany}

\begin{abstract}  A multi-component formalism is developed to describe
three-body systems with nonstatic pairwise interactions and  non-nucleonic
degrees of freedom. The dressed-bag model for $NN$ interaction based on the
formation of an intermediate six-quark bag dressed by a $\sigma$-field is
applied to the $3N$ system, where it results in a new three-body force between
the six-quark bag and a third nucleon. Concise variational calculations of $3N$
bound states are carried out in the dressed-bag model including the new
three-body force. It is shown that this three-body force gives at least half
the $3N$ total binding energy, while the weight of non-nucleonic components in
the $^3$H and $^3$He wavefunctions can exceed 10\%. The new force model
provides a very good description of $3N$ bound states with a reasonable
magnitude of the $\sigma NN$ coupling constant. The model can serve as a
natural bridge between dynamical description of few-nucleon systems and the
very successful Walecka approach to heavy nuclei and nuclear matter. 
\end{abstract}



\section{Introduction}  
In recent years, few-nucleon theory based on
traditional models for the $NN$ force had significant success~\cite{Gl1,FB}. 
In particular, a high level of precision for the 
agreement between $3N$ and $4N$ calculations and the corresponding experimental
data has been reached in many cases. Such a degree of agreement obtained
without any free parameter could  promise a description of all properties of
few-nucleon systems in the near future by solving the exact Faddeev-Yakubovsky
equations with the inclusion of modern $NN$ potentials and $3N$ forces.

However, rather significant disagreements between results of the most exact
few-nucleon calculations and experimental data still remain. Among all such
discrepancies, we mention here only the most important ones such as the $A_y$
puzzle in $\vec{N}+d$ and $\vec{N}+{}^3{\rm He}$ scattering~\cite{FB,Ay},
disagreements at the minima of differential cross sections (Sagara puzzle) at
$E\sim 150 \div 200$~MeV and polarization data for $N+\vec{d}$~\cite{Sagara},
$\vec{N}+d$, $\vec{N}+\vec{d}$~\cite{polnd} and $\vec{N}+{}^3{\rm He}$
scattering.  The strongest discrepancy between current theories and experiments
has been found in studies of the short-range $NN$ correlations in the $^3{\rm
He}(e,e'pp)$~\cite{Nikef}, $^4{\rm He}(\gamma,pp)$~\cite{Grab}, and $^3{\rm
He}(e,e'NN)$~\cite{Jef} processes. The strong short-range correlations found 
in these experiments are likely due to (at least partially) a three-body force
acting in a space region where two nucleons  are close to each other.
Therefore, in order to quantitatively describe such correlations, some new
types of three-body forces should be introduced.

In addition to these particular problems, there are more fundamental problems
in the current theory of nuclear forces, e.g., strong discrepancies between the
$\pi NN$, $\pi N\Delta$ and $\rho NN$ form factors used in OBE models for the
description of elastic and inelastic scattering and in the parameterization of
$2N$ and $3N$ forces~\cite{KuJPG,Gibson,Gibson88,Sauer}. Many of these difficulties are attributed to
a rather poor knowledge of the short-range behaviour of nuclear forces. This
behaviour was traditionally associated with the vector $\omega$-meson exchange.
However, the characteristic range of this $\omega$-exchange (for
$m_{\omega}\simeq 780$~MeV) is equal to about $\lambda_{\omega} \simeq 0.2 \div
0.3$~fm, i.e., is deeply inside the internucleon overlap region. Therefore, the
quark structure of nucleons should be important at this short range. Within
six-quark dynamics in its turn, it has long been
known~\cite{Kus91,Myhrer,Fae83,Progr92,Yam86} that the mixing of the completely
symmetric $s^6[6]$ component with the mixed-symmetry $s^4p^2[42]$ component can
determine the structure of the whole short-range interaction (in the 
$S$-wave). Assuming a reasonable $qq$ interaction model, many authors (see
e.g.~\cite{Oka83,Fujiwara,Stancu,Bartz}) have demonstrated that this  mixture
can result in both strong short-range repulsion (associated mainly with the
$s^6$ component) and intermediate-range attraction (associated mainly with the
above mixed-symmetry $s^4p^2$ component). However, recent
studies~\cite{Stancu,Bartz} for $NN$ scattering on the basis of the newly
developed  Goldstone-boson-exchange (GBE) $qq$ interaction have resulted in  a
purely {\em repulsive} $NN$ contributions from both $s^6[6]$ and $s^4p^2[42]$
six-quark components. There is no need to say that any quark-motivated model
for the $NN$ force with $\pi$-exchange between quarks inevitably leads to the
well-established Yukawa $\pi$-exchange interaction between nucleons at long
distances.

Trying to solve the above problems (and to understand more deeply the mechanism
for the short-range $NN$ interaction), we suggested to add to the
conventional Yukawa meson-exchange ($t$-channel) mechanism (at intermediate and
short ranges) contributions
of $s$-channel graphs describing the formation of a dressed six-quark bag in
an intermediate state such as $|s^6 +\sigma\rangle$ or
$|s^6+2\pi\rangle$\cite{KuJPG,KuInt}. It has been shown that, due to the change
in the symmetry of the six-quark state in the transition from the $NN$ channel
to the intermediate dressed-bag state, the strong scalar $\sigma$-field arises
around the symmetric $6q$ bag. This intensive $\sigma$-field squeezes the bag
and increases its density. The high quark density in the symmetric $6q$ bag
enhances meson field fluctuations around the bag and thereby partially restores
the chiral symmetry. Therefore, the masses of constituent quarks and $\sigma$
mesons decrease. As a result of this phase transition, the dressed bag mass
decreases considerably (i.e., a large gain in energy arises), which manifests
itself as a strong effective attraction in the $NN$ channel at intermediate
distances. This attraction can be described in terms of the OBE model as the
effective $t$-channel $\sigma$-exchange.

However, more accurate calculations of the intermediate-range $NN$ 
interaction~\cite{twopi} within the $2\pi$-exchange
model with the $\pi\pi$ $s$-wave interaction have revealed that this
$t$-channel mechanism cannot give a strong intermediate-range attraction in the
$NN$ sector, which is necessary for binding of a deuteron and fitting of $NN$
phase shifts. The fact that the conventional $t$-channel $2\pi$-exchange with
the $\pi\pi$ $S$-wave interaction and reasonably soft $\Lambda_{\pi NN}$
cut-off values cannot give the sufficiently strong intermediate-range $NN$
attraction has also been corroborated by recent independent
calculations~\cite{Oset}. Thus, the $t$-channel mechanism of the $\sigma$
exchange should be replaced by the corresponding $s$-channel mechanism. The
contribution of the $s$-channel mechanism would generally be much larger due to
resonance-like enhancement\footnote{
$t$-channel mechanism can be associated with the direct nuclear reaction where
only a few degrees of freedom are important, while the  $s$-channel mechanism
can be associated with resonance-like (or compound nucleus like) nuclear
reactions with much larger cross sections at low energies.}

Based on this $s$-channel mechanism, we proposed a new model for the $NN$
interaction (referred to as the ``dressed bag model'' (DBM))~\cite{KuJPG,KuInt}
that provides a quite good description of both $NN$ phase  shifts up to 1~GeV
and the deuteron structure. The developed model includes both the conventional
$t$-channel contributions (Yukawa $\pi$ and $2\pi$-exchanges) at long and
intermediate distances and the $s$-channel contributions due to the formation
of intermediate dressed-bag states at short distances. The most important
distinction of such an approach from conventional models for nuclear forces is
the explicit presence of a non-nucleonic component in the total wavefunction of
the system, which necessarily implies the presence of new three-body forces
(3BF) of several kinds in the $3N$ system. These new three-body forces differ
from conventionally used models for three-body forces. One important aspect of
the three-body force should be emphasized here. In conventional OBE models, the
main contribution to $NN$ attraction is due to the $t$-channel $\sigma$
exchange. However, the three-body force models suggested until now (such as
Urbana-Illinois or Tucson-Melbourne) are mainly based on the two-pion exchange
with intermediate $\Delta$-isobar production, and the $\sigma$-exchange either
is not taken into account or is of little importance in these models. In
contrast, $\sigma$-exchange in our approach dominates in both $NN$  and $3N$
forces; i.e., the general pattern of the nuclear interaction appears to be more
consistent.

The aim of this work is just to study both the new type of three-body forces
and the properties of the $3N$ system with the dressed-bag model (DBM) proposed
for $NN$ and $3N$ forces. In particular, we will show that the role of 3BF
changes remarkably compared to the conventional OBE models. In the conventional
models, the main contribution to the $3N$ binding energy comes from the pair
$NN$ force, whereas 3BF gives a certain (although significant) correction of
about 15\% to the total binding energy. In contrast, in our DBM approach the
contribution of the new scalar three-body forces turns out to be larger by
several times and is equal to about half the total $3N$ binding energy.

This paper is organized as follows. In Sect.II, we give a brief description of
the DBM for the $NN$ system. In Sect.~III, we develop the multi-component
formalism for the $3N$ system. In Sect.~IV, the general formalism is specified
to the one-pole approximation for the dressed-bag resolvent, which is used in
our calculations. In Sect.~V, three types of 3BF are considered: one-meson
exchange ($\pi$ and $\sigma$) between the dressed bag and third nucleon and the
$2\sigma$-exchange with breaking of the $\sigma$-loop in the two-nucleon
interaction. The results of our variational calculation of the $3N$ system are
given in Sect.VI and the results are discussed in Sect.VII. In the Conclusion
we summarize the main results of the work.

\section{Dressed bag model for $NN$ forces}

Here, we give a brief description of the two-component dressed-bag model for
the $NN$ interaction. The detailed description has been presented in our
previous papers~\cite{KuJPG,KuInt}. We consider a system that can exist in two
different phase states: the $NN$ phase and dressed six-quark bag phase. In the
$NN$ channel, the system can be described as two nucleons interacting via
one-boson exchange; in the $6q+\sigma$ channel, the system is treated as a
six-quark bag surrounded by the strong scalar-isoscalar $\sigma$-field (a
"dressed" bag).\footnote{%
Full description of the $NN$ interaction at energies
$E\sim 1$~GeV still requires other fields in the bag such as $2\pi$, $\rho$ and
$\omega$ but here we employ the version of DBM including only
 a leading $\sigma$-field~\cite{KuInt} with parameters which are determined
 either phenomenologically from fit of $NN$ phase shifts (variant (I)) or 
taken directly from the respective $\sigma$-loop diagram (the variant (II)).} 
Accordingly, the wavefunction of the system consists of two
components of different nature:
 \begin{equation}
\Psi=\left ( \begin{array}{l} \Psi_{NN} \\ \Psi_{6q+\sigma}\end{array}
\right ).
 \label{Fock}
\end{equation}
 The Hamiltonian of the system has the corresponding matrix form
\begin{equation}
\hat{H}=\left (
\begin{array}{ll} H_{NN} & H_{NN,6q\!+\!\sigma}\\
 H_{6q\!+\!\sigma ,NN} & H_{6q\!+\!\sigma} \end{array}
 \right ).
\label{Ham}
\end{equation}
 Here, $H_{NN}$ includes the direct $NN$ interaction induced by one- and
two-pion exchanges with the $\pi NN$ form factors taken with the correct
``soft'' cut-off parameters~\cite{KuJPG,KuInt}. Thus, $H_{NN}$ describes the
peripheral part (at $r_{NN} > 1$~fm) of the $NN$ interaction.
$H_{6q\!+\!\sigma}$ is the Hamiltonian of the dressed bag, while the operators
$H_{6q\!+\!\sigma,NN}$ and $H_{NN,6q\!+\!\sigma}$ describe the transitions from
$NN$ to the $(6q+\sigma)$ channel and vice versa. The Schr\"odinger equation%
\footnote{In the force models described here the $s$-channel mesons are treated
relativistically, while the heavy $6q$ bag and nucleons are taken in the
nonrelativistic approximation. This is quite reasonable at the energy range
$E_{\rm Lab}\le 0.8$~GeV.} 
reduced to two coupled equations and, by excluding the second component
$\Psi_{6q\!+\!\sigma}$, one obtains the following equation for the proper $NN$
channel wavefunction with the effective Hamiltonian:
 \begin{equation}
  H_{NN}\Psi_{NN}+  H_{NN,6q\!+\!\sigma}(E-H_{6q\!+\!\sigma})^{-1}
  H_{6q\!+\!\sigma ,NN}
  \Psi_{NN}= E\,\Psi_{NN}.
\label{eff1}
\end{equation}
Having obtained the solution of this equation for the $\Psi_{NN}$ component, one can
also find straightforwardly the second component $\Psi_{6q\!+\!\sigma}$:
 \begin{equation}
 \Psi_{6q\!+\!\sigma}=(E-H_{6q\!+\!\sigma})^{-1} H_{6q\!+\!\sigma ,NN}\Psi_{NN}
\label{2comp}
\end{equation}

Using the following simple one-pole approximation for the dressed bag 
resolvent $(E-H_{6q\!+\!\sigma})^{-1}$:
 \begin{equation}
(E-H_{6q\!+\!\sigma})^{-1}=\sum_{\alpha}\int\frac{|\alpha,{\bf k}\rangle
\langle \alpha,{\bf k}|}{E-E_{6q\!+\!\sigma}(\alpha ,{\bf k})}{\rm d}{\bf k},
\label{resn}
\end{equation}
where $|\alpha\rangle$ is the $6q$ part of the wavefunction for the dressed bag
and $|{\bf k}\rangle$ represents the plane wave of the  $\sigma$-meson
propagation, one obtains an effective interaction in the NN channel as a sum of
separable terms.\footnote{%
The single-pole approximation for the dressed-bag resolvent should  be quite
valid at incident energies in $NN$-channel $E_{\rm lab} < 1$~GeV  because  the
energy spacing $\hbar\omega$  between different excited states  of the dressed
bag (with different parity and number of radial nodes)  is expected  around
300-350~MeV. So that the spacings between excited states in the  fixed channel 
(with the same parity) should be around $2\hbar\omega =600\div 700$~MeV in the
c.m. system. Hence a  few-quantum excitations of the bare $6q$ bag should lie
very high in the spectrum.} 
We emphasize that the sign of the dressed bag
wavefunction $\Psi_{6q+\sigma}$ at low and intermediate energies $E<E_0$ (where
$E_0$ is the lowest eigenenergy of the dressed bag) is {\em opposite} to that
of the $NN$ wavefunction.

To solve equation (\ref{eff1}) in such an approximation, the knowledge of 
total Hamiltonian
$H_{6q\!+\!\sigma}$ of the dressed bag, as well as the total transition
operator $H_{NN,6q\!+\!\sigma}$, is not necessary. Only the projections of the
transition operator $H_{NN,6q\!+\!\sigma}|\alpha \rangle$ onto the bag states
$|\alpha \rangle $ are necessary to construct the interaction in the $NN$
channel. These projections have been calculated in the microscopic quark-meson 
model~\cite{KuJPG,KuInt}. The effective interaction $V_{NqN}$ induced in
the $NN$ channel by coupling with the intermediate dressed bag state is
illustrated by the graph in Fig.~1.

\begin{figure}[h]
\begin{center}
 \epsfig{file=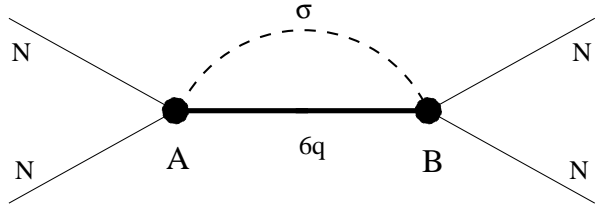,width=0.5\textwidth}
\end{center}

\caption{ Effective $NN$ interaction induced by the formation of an
intermediate dressed bag}
\end{figure}

In general, the state index $\alpha$ includes all quantum numbers of the
dressed bag, i.e. $  \alpha\equiv \{J,M,S,T,L_b,L_{\sigma}\}$, where $L_b$,
$S$, $T$, $J$, and $M$ are the orbital angular momentum of the $6q$ bag, its
spin, isospin, total angular momentum, and its projection on the $z$ axis,
respectively, and $L_{\sigma}$ is the orbital angular momentum of the $\sigma$
meson. However, in the present version of the DBM, only the s-wave state of the
$6q$ bag with the $s^6$ configuration is taken into account, so that $L_b=0$,
$J=S$, and the isospin of the bag is uniquely determined by its spin. The
states of the dressed bag with $L_{\sigma}\ne 0 $ should lie higher than those
with $L_{\sigma} =0$. For this reason, the former states are not included in
the present version of the model. Therefore, the state index $\alpha$ is
specified here by the total angular momentum of the bag $J$ and (if necessary)
its $z$ projection $M$: $ \alpha \Rightarrow  \{J(M)\}$.

Thus, the effective interaction in the $NN$ channel
$V_{NqN}\equiv H_{NN,6q\!+\!\sigma}(E-H_{6q\!+\!\sigma})^{-1}
  H_{6q\!+\!\sigma ,NN}$, after partial-wave decomposition,
can be written as a sum of separable terms in each partial wave:
 \begin{equation}
V_{NqN}=\sum_{J ,L,L'}V^{J}_{LL'}({\bf r},{\bf r}^{\prime}),
 \label{nqn2}
 \end{equation}
 with
\begin{equation}
V^{J}_{LL^{\prime}}({\bf r},{\bf r'})= \sum_M \varphi^{JM}_{L}({\bf
r})\,\lambda^{J}_{LL^{\prime}}(E)\, {\varphi^{JM}_{L^{\prime}}}^*({\bf r'}).
 \label{zlz}
 \end{equation}
The energy-dependent coupling constants $\lambda^{J}_{LL^{\prime}}(E)$
appearing in eq. (\ref{zlz}) are calculated from the diagram shown in Fig.~1;
i.e., they are expressed in terms of the loop integral of the product of two
transition vertices $B$ and the convolution of two  propagators for the meson
and quark bag with respect to the momentum $k$:
 \begin{equation}
\lambda^{J}_{LL^{\prime}}(E)=\int^{\infty}_0\,{\rm d}{\bf k}
\frac{B_L^J({\bf k},E)\,{B_{L^{\prime}}^J}^*({\bf k},E)}
{E-m_{\alpha}-\varepsilon_{\sigma}(k)},
 \label{lamb}
 \end{equation}
 where $\varepsilon_{\sigma}(k)$ is the kinetic energy of the dressed bag:
 $\varepsilon_{\sigma}(k)= k^2/2m_{\alpha}+\omega_{\sigma}(k)
 \simeq m_{\sigma}+k^2/2\bar{m}_{\sigma}$, $\bar{m}_{\sigma}$ is the
 reduced mass in the $6q+\sigma$ channel
 $\bar{m}_{\sigma}=m_{\sigma}m_{\alpha}/(m_{\sigma}+m_{\alpha})^{-1}$,
 and  $\omega_{\sigma}(k)$ being relativistic energy of $\sigma$-meson:
 $\omega_{\sigma}(k)=\sqrt{m_{\sigma}^2+k^2}$.
The vertex form factors $B^{J}_{L}(k)$ and the potential form factors
$\varphi^{JM}_{L}({\bf r})$
 have been calculated in the microscopic quark-meson
model~\cite{KuJPG,KuInt}.

 When the $NN$-channel wavefunction $\Psi_{NN}$ is obtained by solving the
Schr\"odinger equation with the effective Hamiltonian $H_{NN} + V_{NqN}$, the
$6qN$-component of the wavefunction can be found algebraically from
eq.~(\ref{2comp}):
 \begin{equation}
|\Psi^{JM}_{6q\!+\!\sigma}(E)\rangle =
\Psi_{\sigma}^{J}({\bf k},E)\,|\alpha^{JM}\rangle ,
\label{psi6q}
 \end{equation}
where the function
 \begin{equation}
  \Psi_{\sigma}^{J}({\bf k},E)
= \sum_{L} C_{L}^{J}(E)
\frac{B_L^J({\bf k},E)}
{E-m_{\alpha}-\varepsilon_{\sigma}(k)}
\label{psisig}
\end{equation}
 can be interpreted as the mesonic part of the dressed-bag wavefunction. The
coefficients $C_{L}^{J}(E)$ in eq.~(\ref{psisig}) are overlap integrals of the
nucleon-nucleon wavefunction and the form factors $\varphi^{JM}_{L}({\bf r})$
of the separable  potential (\ref{zlz}):
\begin{equation}
C_{L}^{J}(E)=\int \Psi_{NN}({\bf r},E)\,\varphi^J_{L}({\bf r}) {\rm d}{\bf r}.
\label{CLJ}
\end{equation}

The contribution of the dressed-bag component to the total wavefunction 
is proportional to the
norm of $\Psi_{6q\!+\!\sigma}$: 
\begin{eqnarray} 
\|\Psi_{6q\!+\!\sigma}\|^2 = \|\alpha\|^2 \int
\Psi^{JM}_{\sigma}({\bf k},E)\Psi^{JM*}_{\sigma}({\bf k},E) 
{\rm d}{\bf k} \nonumber \\
 =\sum_{LL'}
C_{L}^{J}(E)\,{C_L'^{J}}^*(E) \int_0^{\infty} \frac{B_L^J(k,E)\,
{B_{L'}^J}^*(k,E)}
{(E-m_{\alpha}-\varepsilon_{\sigma}(k))^2}{\rm d}{\bf k} \nonumber\\
 = \sum_{LL'}C_{L}^{J}(E)\,{C_{L'}^{J}}^*(E)P_{LL'}^{J}(E).
\label{normsig}
\end{eqnarray}
 If the vertex form factors ${B_L^J}$ are  energy independent, 
 the factors $P_{LL'}^{J}(E)$ are simply related to
 the coupling constants $\lambda^{J}_{LL^{\prime}}(E)$ (see eq.(\ref{lamb})):
  \begin{equation}
P_{LL'}^{J}(E)=-\,\frac{{\rm d}\lambda^{J}_{LL'}(E)}{{\rm d}E}.
 \label{deriv}
 \end{equation}

The total wavefunction of the bound state $\Psi$ must be normalized.
Assuming that the nucleonic part of the wavefunction $\Psi_{NN}$
found from the effective Schr\"odinger equation has the standard normalization
 $\|\Psi_{NN}\|=1 $,
 one obtains that the weight of the dressed bag component is equal to
\begin{equation}
P_{6q}=\frac{\|\Psi_{6q\!+\!\sigma}\|^2}{(1+\|\Psi_{6q\!+\!\sigma}\|^2)}.
\label{beta2}
\end{equation}

\begin{table}[h]
\caption {Deuteron properties in DBM.}
\medskip

\begin{tabular}{|c|c|c|c|c|c|c|c|}\hline &&&&&&&\\
Model    & $E_d$(MeV)& $P_D$(\%)& $r_m$(fm) & $Q_d$(fm$^2$) &
$\mu_d\,(\mu_N)$ & $A_S$(fm$^{-\!1/2}$)& $\eta (D/S)$ \\ &&&&&&&\\ \hline
 RSC      &
2.22461  & 6.47 & 1.957 & 0.2796 & 0.8429 & 0.8776 & 0.0262  \\ \hline

 Moscow 99   &2.22452  & 5.52 & 1.966 & 0.2722 & 0.8483 & 0.8844 & 0.0255 \\
  \hline
 Bonn 2001 &2.224575 & 4.85 & 1.966 & 0.270  & 0.8521 & 0.8846 & 0.0256 \\
  \hline
 { DBM (var. a) } &2.22454 & 5.22&1.9715&0.2754& 0.8548&0.8864& 0.0259\\
 $P_{6q}=3.66\%
 $ &&&&&&&\\ \hline
 { DBM (var. b) } &2.22459 & 5.31&1.970&0.2768& 0.8538&0.8866& 0.0263\\
 $P_{6q}=2.5 \%
 $ &&&&&&&\\ \hline

experiment&2.224575 &        & 1.971 & 0.2859 & 0.8574 & 0.8846 & 0.0263$^*$ \\
 \hline
\end{tabular}

\noindent {\small $^*$ An average value of the asymptotic mixing parameter 
$\eta$ over a few most accurate results is presented here 
(see refs.~\cite{Hor,Berth,Must,Rodn}).}

\end{table}

The model constructed above gives a very good description for the
coupled-channel $^3S_1- {}^3D_1$ phase shifts, mixing parameter $\varepsilon_1$
and singlet $^1S_0$ shifts in the energy region from zero up to 1
GeV~\cite{KuInt}. The deuteron observables obtained in this model without any
additional parameter are presented in Table~1 in comparison with some other
$NN$ models and experimental values. The quality of agreement with experimental
data for the deuteron static properties found with the presented force model,
in general, is higher than those for the modern $NN$ potential model such as
Bonn, Argonne, etc., especially for the asymptotic mixing parameter $\eta$ and
the deuteron magnetic and quadrupole moments. The weight of the dressed bag 
component in the deuteron is calculated from the energy dependence of the 
coupling constants $\lambda^J_{LL'}$ (see eqs.~(\ref{deriv}) and (\ref{beta2}) 
and is varied from 2.5\% to 3.6\% in different versions of the 
model~\cite{KuJPG,KuInt}.

\section{DBM for the three-nucleon system}
When generalizing the above model to the $3N$ system, one meets two
difficulties: the appearance of the energy dependent coupling constants
$\lambda^J_{LL'}(E)$ and the presence of non-nucleonic degrees of freedom in
{\em an explicit form}. Thus, we suggest that the $3N$ system can be found in
the five different states (in three phases): $NNN$,
$N_1\!+\!(6q^{(23)}\sigma)$, $N_2+(6q^{(31)}\sigma)$, $N_3+(6q^{(12)}\sigma)$,
and $9q+\sigma$. Therefore, the total wave function can be represented as the
five-row Fock column
\begin{equation}
\Psi=\left ( \begin{array}{l} \Psi_{3N} \\
  \Psi^{(1)}_{6qN} \\
  \Psi^{(2)}_{6qN} \\
  \Psi^{(3)}_{6qN} \\
  \Psi_{9q+\sigma}
\end{array}\right ).
\label{Fock3}
\end{equation}

The nucleonic component $\Psi_{3N} $ here describes the system as three
nucleons interacting at large distances via (one- and two-) boson exchanges. In
the $(6qN)^{(i)}$ channels, there are the dressed $6q$ bag and $i$th nucleon
($i=1,2,3$). Hereinafter, the symbol $\sigma$ in the notation of these channels
is omitted for brevity. The wavefunction component $\Psi^{(i)}_{6qN}$ depends
on the dressed bag variables, including the momentum of the $\sigma$-meson, and
the $6q\!-\!N$ relative motion variable (the Jacobi coordinate or momentum).
The fifth component of the total wavefunction (\ref{Fock3}) describes a dressed
nine-quark bag.   

  The admixture of the dressed $9q$-bag in $3N$ system  and in "normal" nuclei
should be rather low, at least one order of magnitude lower as compared to the
weight of the $6q$ bag. This can be easily explained  by the fact the
$9q$-clusters in nuclei can  be only generated from $6q$ and $3q$-clusters at
their fusion which has rather  low probability due to very high kinetic energy
of $9q$ system. So the weight  of $9q$ configuration must be a small fraction
of that for $6q+3q$ component. Thus one can expect only a  very minor effects
of this $9q$ configurations except the region of very  high momentum transfer.


Therefore, we neglect this component in present study and consider below the
4-component system with the (4x4)-matrix Hamiltonian
\begin{equation}
\hat{H}=\left ( \begin{array}{llll}
  H_{3N} & H_{3N,1} & H_{3N,2} & H_{3N,3}  \\
  H_{1,3N} & H_1 &    0 & 0 \\
  H_{2,3N} & 0   &  H_2 & 0  \\
  H_{3,3N} & 0   &    0 &H_3  \\
\end{array}\right ).
\label{Ham3}
\end{equation}
 Here, we use the brief notation for the following parts of the Hamiltonian: the
operator  $H_{3N,i} \equiv  H_{3N,(6q^{(jk)}\!+\!\sigma)+N_i}$ describes the
transition from the $\{(6q^{(jk)}\!+\!\sigma)+N_i\}$ channel to the pure
three-nucleon channel; the Hamiltonian $H_i \equiv H_{6q\!+\!N_i}$ describes
the interacting system of the dressed $6q$ bag and $i$th nucleon. 

In our $3N$ dynamics, we neglect the direct transitions between different 
$6q-N$ channels. Such channels are coupled only via an intermediate 
$3N$ configurations. This assumption looks absolutely natural, while  
the incorporation of direct transitions between such channels 
 requires some exotic new mechanisms for the coupling.

The Hamiltonian $H_{3N}$ of the nucleonic channel includes the kinetic energy
of the relative motion of three nucleons and the direct $NN$ interactions (OPE
and TPE with soft cut-offs) that enter into the two-body Hamiltonian:
 \begin{equation}
h_{NN}= t + V_{NN} \qquad \Rightarrow \qquad H_{3N}=T+\sum_{i=1}^3 V_{NN}^{(i)}.
\label{H3N}
\end{equation}
Hereinafter, we use lowercase letters ($g$, $h$, etc.) for two-body
operators and capital letters for three-body ones.

The ($6qN \to 3N$)-transition operators $H_{3N,i}$ do not affect the third 
nucleon and therefore take the form:
\begin{equation}
H_{3N,i}= h^{(jk)}_{NN,6q\!+\!\sigma}\otimes {\bf 1}^{(i)},
\mbox{ where }
(ijk)=(123),(231),(312),
\label{Hi3N}
\end{equation}
 and $h^{(jk)}_{NN,6q\!+\!\sigma}$ are the two-body DBM transition operators 
 for pair ($jk$).

Since direct transitions between different $6qN$ channels are absent,
 the exclusion of the $6qN$ components is completely similar to the
two-body case: hence one can rewrite the 4-component Hamiltonian
in the two-component form
\begin{equation}
\hat{H}=\left (
\begin{array}{ll} H_{3N} & H_{3N,6qN}\\
 H_{6qN,3N} & H_{6qN} \end{array}
 \right ),
\label{Ham32}
\end{equation}
 where $H_{6qN}$ is the {\em diagonal} (3x3)-matrix operator with the elements
$H_i$. The resolvent of $H_{6qN}$ is also a diagonal (3x3)-matrix.

 The Hamiltonian $H_i$ describes the dressed bag $6q(jk)$ (formed by the fusion
of the $j$th and $k$th nucleons) interacting as a whole with the third ($i$th) 
nucleon via exchanges by mesons ($\pi$, $\sigma$). Just this interaction
results in the new three-body force arising in the two-phase model. The
Hamiltonian
 \begin{equation}
H_i= h_{6q\!+\!\sigma}(jk) \oplus h_{6q,N^{(i)}},
\label{Hi}
\end{equation}
is the direct sum of the dressed-bag Hamiltonian $h_{6q\!+\!\sigma}(jk)$ and
single-particle Hamiltonian of the $i$th nucleon $h_{6q,N^{(i)}}$, which
includes its kinetic energy  and the potential energy of its interaction with
the bag:
 \begin{equation}
   h_{6q,N^{(i)}} = t_i + V_{6q,N^{(i)}}
\label{H6qN}
\end{equation}
It depends only on the single Jacobi coordinate (or momentum) of the third
($i$th) nucleon. Therefore, the resolvent of $H_i$ is equal to the 
convolution of two
subresolvents: the first for the dressed-bag Hamiltonian
$(E-h_{6q\!+\!\sigma}(jk))^{-1}$ and the second for the single-particle
Hamiltonian (\ref{H6qN}):
 \begin{equation}
(E-H_i)^{-1}=-\frac{1}{2\pi\rm i}\int_{-\infty}^{\infty}
(\varepsilon-h_{6q\!+\!\sigma}(jk))^{-1}\, 
{\cal G}_{6q,N_i}(E-\varepsilon)\, d\varepsilon,
\label{resi}
\end{equation}
where
  \begin{equation}
  {\cal G}_{6q,N_i}(E)= (E-h_{6q,N^{(i)}})^{-1}.
\label{g6qN}
\end{equation}

Excluding the $6qN$ channels from the Schr\"odinger equation for the total
four-component wave function, one arrives at the following effective
Hamiltonian  ${\cal H}_{3N}$ acting in the $3N$ channel:
 \begin{equation}
   {\cal H}_{3N} = H_{3N} + \sum_{i=1}^3 H_{3N,i}(E-H_i)^{-1}H_{i,3N}
\label{H3eff}
\end{equation}
which has the form of the standard three-body Hamiltonian with pairwise {\em
energy-dependent} interactions:
 \begin{equation}
   {\cal H}_{3N} = T+\sum_{i=1}^3(V_{NN}^{(i)} +V_{3N,6qN,3N}^{(i)});
\label{H3veff}
\end{equation}

This form of the effective Hamiltonian is quite suitable for the standard
Faddeev reduction. Although the effective interaction is the sum of three
"pairwise" interactions, each term $V_{3N,6qN,3N}^{(i)}$ describes not a
genuine two-body force but a three-body force including the interaction
between the $6q$ bag and third nucleon. Moreover, even if this interaction is
disregarded, the potential $V_{3N,6qN,3N}^{(i)}$ still depends on the momentum
of the third nucleon. This dependence on the momentum of the third particle
reduces the strength of the effective interaction between two other particles
due to a specific energy dependence of the coupling constants.
Actually, the form of the effective Hamiltonian (\ref{H3eff}) implies that
there are no pure two-body forces in the $3N$ system in  our approach, except
the peripheral part of the OBE interaction.

 After solving the Faddeev equations (or the Schr\"odinger equation for bound
states) with the effective Hamiltonian (\ref{H3eff}) for the $3N$ wavefunction,
the "bag" components of the total wave function are ``recovered'' by means of
the relationship:
 \begin{equation}
   \Psi^{(i)}_{6qN} = (E-H_i)^{-1}H_{i,3N}\Psi_{3N}.
\label{psi6qN}
\end{equation}

\section{Single-pole approximation for the dressed bag resolvent}
 The momentum representation is more appropriate for description of the $3N$ 
 system in the case of DBM. We will
 employ the same notation for functions both in the coordinate and momentum
 representations. In the single-pole approximation for the dressed-bag
 resolvent, the resolvent $(E-H_i)^{-1}$ (\ref{resi}) can be reduced to a sum
 of factorized terms:
 \begin{equation}
\fl (E-H_i)^{-1}=\sum_{\alpha} \int|\alpha^{(i)} ,{\bf k}\rangle\,
{\cal G}_{6q,N_i}(E\!-\!m_{\alpha}\!-\!\varepsilon_{\sigma}(k))\,
\langle \alpha^{(i)},{\bf k}|\,{\rm d}{\bf k}
\label{resi1}
\end{equation}
 and the effective $3N$ interaction $V_{3N,6qN,3N}^{(i)} $ becomes the sum
of integral operators with factorized kernels:
 \begin{equation}
\fl V_{3N,6qN,3N}^{(i)}({\bf p_i, p'_i, q_i,q'_i};E)=
 \sum_{JM,J'M',L,L'}\varphi^{JM}_L({\bf p_i})W^{JJ'}_{LL'}({\bf q_i,q'_i};E)
 \varphi^{J'M'}_{L'}({\bf p'_i}),
 \label{kernel}
\end{equation}
where $W^{JJ'}_{LL'}$ is expressed in terms of the integral of the product 
of the vertex functions and one-particle resolvent (\ref{g6qN}):
 \begin{equation}
 W^{JJ'}_{LL'}({\bf q_i,q'_i};E)=\int {\rm d}{\bf k} B^J_L({\bf k})\,
 {\cal G}_{6q,N_i}({\bf q_i,q'_i};E\!-\!{m_{\alpha}}\!-
 \!\varepsilon_{\sigma}(k)\!)\, B^{J'}_{L'}({\bf k}).
  \label{W}
\end{equation}
If the interaction between the $6q$ bag and third nucleon is neglected, 
the resolvent ${\cal G}_{6q,N}$ becomes the simple propagator:
 \begin{equation}
\fl  {\cal G}_{6q,N}({\bf q_i,q'_i};E\!-\!{m_{\alpha}}\!-\!\varepsilon_{\sigma}(k)\!)
\Rightarrow \delta({\bf q_i-q'_i})
\frac{1}{E\!-\!{m_{\alpha}}\!-\!\varepsilon_{\sigma}(k)\!-\!q^2/2m}; 
\;m=\frac{m_Nm_{\alpha}}{m_N+m_{\alpha}}.
  \label{g0}
\end{equation}
Then, the effective interaction reduces to the sum of two-body separable
potentials with the coupling  constants depending on the total three-body
energy $E$ and third-particle momentum $q_i$:
 \begin{equation}
 \fl \tilde{V}_{3N,6qN,3N}^{(i)}({\bf p_i, p'_i, q_i,q'_i};E)= 
  \delta({\bf q_i-q'_i})
 \sum_{J,M,L,L'}\varphi^{JM}_L({\bf p_i})\,\lambda^{J}_{LL'}
 (E-\frac{q_i^2}{2m})\,\varphi^{JM}_{L'}({\bf p'_i}),
 \label{kernel0}
\end{equation}
where the coupling constants $\lambda^{J}_{LL'}$ are determined by
eq.(\ref{lamb}) with replacing $E$ by $E-\frac{q_i^2}{2m}$. In other words,
the effective interaction in the three-body system reduces to a
sum of three effective pairwise interactions $V_{NqN}$ depending on the
{\em total three-body energy} and momentum of the third (spectator)
particle:
 \begin{equation}
 V_{3N,6qN,3N}\Rightarrow 
 \sum_{i=1}^3 \delta({\bf q}_i\!-\!{\bf q'}_i)\,V_{NqN}^{(i)}
 ({\bf p_i,p'_i};E\!-\!\frac{q_i^2}{2m}).
 \label{3NqN}
\end{equation}
It should be noted that this result does not depend on specific choice of
interaction model, but is quite general provided that there are not direct
transition between different $6qN$ channels and the transition operators 
take the form (\ref{Hi3N}) 

\begin{figure}[h]
\begin{center}
{\epsfig{file=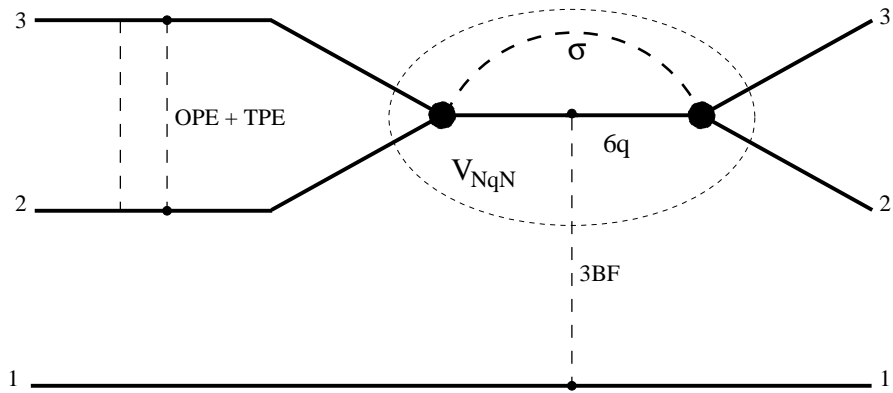,width=0.75\textwidth}}

\caption{Different interactions in the $3N$ system for one of
three possible combinations (1+23): the peripheral two-nucleon interaction is 
due to OPE $+$ TPE, effective two-body interaction $V_{NqN}^{(1)}$ is induced
by the formation of dressed six-quark bag and meson-exchange three-body
force (3BF).}
\end{center}
\end{figure}

When using the effective interaction (\ref{3NqN}), one must also include an
additional three-body force due to the meson-exchange interaction between the
dressed bag and third nucleon. The pattern of different interactions arising in
the $3N$ system in such a way is illustrated in Fig.~2.

In the single-pole approximation, the components of the "bag" wave  function 
(\ref{psi6qN}) can be expressed in terms of the nucleonic component as
 \begin{equation}
 \Psi^{(i)}_{6qN}({\bf k,q}_i;E) = \sum_{J_iM_iL_i}|\alpha^{J_iM_i}\rangle
\frac{B_{L_i}^{J_i}({\bf k})\,\chi_{L_i}^{J_iM_i}({\bf q}_i)}
{E-m_{\alpha}-\frac{q_i^2}{2m}-\varepsilon_{\sigma}(k)},
 \label{psi6qNIJ}
 \end{equation}
where $\chi_{L_i}^{J_iM_i}({\bf q}_i)$ are the overlap integrals of the 
three-nucleon component of the wavefunction and the potential form factors 
$\varphi^{J_iM_i}_{L_i}$:
 \begin{equation}
 \chi^{J_iM_i}_{L_i}({\bf q}_i) =
 \int \varphi^{J_iM_i}_{L_i}({\bf p}_i) \, \Psi_{3N}({\bf p_i, q_i})\, {\rm
 d}{\bf p_i}.
\label{ovJL}
 \end{equation}
  These overlap functions depend on the momentum (or coordinate), spin, and
isospin of the third nucleon. For brevity, the spin-isospin parts of the
overlap functions and corresponding quantum numbers are omitted unless they are
needed. In eqs.(\ref{psi6qNIJ})-(\ref{ovJL}) and below, we keep the index $i$
in the quantum numbers $L$ and $J$ in order to distinguish the orbital and
total angular momenta in the form factors from the respective angular momenta of
the whole $3N$ system.

It should be noted that the angular part of the function $\chi_{L_i}^{J_i}({\bf
q}_i)$ in eq.(\ref{ovJL}) is not equal to $Y_{L_iM_{L_i}}(\hat{q})$. This part
includes other angular momenta due to coupling between the angular momenta and
spins of the dressed bag and third nucleon. The total overlap function (without
isospin part) can be written, for example, as (we use here the same letter for 
notation of both the total wavefunction and its radial part):
 \begin{equation}
 \chi^{J_iM_i}_{L_i}({\bf q}_i) = \sum_{l_i{\cal J}m_{_{\cal J}}}
 \chi^{J_i}_{L_i,l_i}(q_i)\,
 \langle {\cal J}m_{_{\cal J}}J_iM_i|JM\rangle \,
 {\cal Y}^{{\cal J}m_{\cal J}}_{l_i1/2}({\hat{\bf q}_i}).
\label{ovfull}
 \end{equation}
  Here, $J$ and $M$ are the total angular momentum of the $3N$ system and its
$z$ projection, respectively; $l_i$ and $\cal J$ are the orbital and total
angular momenta of the third ($i$th) nucleon, respectively; $J_i$ is the total
angular momentum of the bag ($J_i=S_i$, because in our case the bag has zero
orbital momentum); and $L_i$ is the orbital angular momentum of a nucleon pair
($jk$) related to the transition vertex $(NN \to 6q\!+\!\sigma)$. For the total
angular momentum $J=1/2$ (the ground state of $^3$H and $^3$He), the orbital
angular momentum of the third nucleon $l_i$ (in the c.m.s. of the entire
system) can take the values $J_i\pm 1$ (we mean here the $6qN$ component only).
Therefore, $l_i$ in our present model can be equal to 0 or 2.

The total norm of all three  $6qN$ components for the $3N$ bound state is 
determined by the integral
 \begin{equation}
 \fl \|\Psi_{6qN}\|^2 = 3\sum_{J_i}\|\alpha^{J_i}\|^2 \sum_{L_iL'_i}
 \int \chi_{L_i}^{J_i}({\bf q}_i)
 \left\{\int \frac{B_{L_i}^{J_i}({\bf k}) B_{L'_i}^{J_i}({\bf k})}
{(E-m_{\alpha}-\frac{q_i^2}{2m}-\varepsilon_{\sigma}(k))^2}
\,{\rm d}{\bf k}\right \}
   \chi_{L'_i}^{J_i}({\bf q}_i)\,{\rm d}{\bf q}_i,
 \label{norm6qN}
 \end{equation}
If the vertex functions $B_{L_i}^{J_i}({\bf k})$ are energy independent
(it was just conjectured when deriving the effective Hamiltonian
(\ref{H3eff})), the internal loop integral with respect to ${\bf k}$ in
eq.(\ref{norm6qN}) (in braces) can be replaced by the energy
derivative of $\lambda^J_{LL'}$: $$
 \int \frac{B_{L_i}^{J_i}({\bf k}) B_{L'_i}^{J_i}({\bf k})\, }
{(E-m_{\alpha}-\frac{q_i^2}{2m}-\varepsilon_{\sigma}(k))^2}\,{\rm d}{\bf k}
=-\frac{\rm d}{{\rm d}E}\, \int \frac{B_{L_i}^{J_i}({\bf k}) B_{L'_i}^{J_i}({\bf k})\, }
{E-m_{\alpha}-\frac{q_i^2}{2m}-\varepsilon_{\sigma}(k)}\,{\rm d}{\bf k}
$$
 \begin{equation}
=-\frac{\rm d}{{\rm d}E}\lambda^{J_i}_{L_iL'_i}(E-\frac{q_i^2}{2m}).
 \label{repln}
 \end{equation}
 Thus, the weight of the $6qN$ component in the $3N$ system is determined by the
same energy dependence of the coupling constants $\lambda^J_{LL'}(\varepsilon)$ as
the contribution of the $6q$ component in the $NN$ system but {\em at a shifted
energy}.

Using eq.(\ref{repln}), the norm of $6qN$ component can be rewritten
eventually as
 \begin{equation}
 \fl \|\Psi_{6qN}\|^2 = 3\,\sum_{J_i}\|\alpha^{J_i}\| \sum_{L_iL'_i}
 \int \chi_{L_i}^{J_i}({\bf q}_i)
 \left (-\frac{\rm d}{{\rm d}E}\lambda^J_L(E\!-\!q_i^2/2m)\right )
  \chi_{L'_i}^{J_i}({\bf q}_i)\,{\rm d}{\bf q}_i,
 \label{norm1}
 \end{equation}
 Due to the explicit presence of the  meson variables in our approach, it is
generally impossible to define the wavefunction of the third nucleon in the
$6qN$ channel. However, by integrating $\Psi_{6qN}({\bf k,q})$ with respect to
the meson momentum $\bf k$, one can obtain the momentum distribution of the
third nucleon in the $6qN$ channel weighted with the $\sigma$-meson momentum
distribution. Based on eq.(\ref{norm1}), we can attribute the meaning of the
wavefunction of the third nucleon in the $6qN$ channel to the quantity

 \begin{equation}
 \tilde{\psi}_{L_i}^{J_iM_i}({\bf q}_i) =
  \sqrt{ \left (-\frac{\rm d}{{\rm d}E}\lambda^{J_i}_{L_iL'_i}(E\!-\!q_i^2/2m) \right )}
  \chi_{L_i}^{J_iM_i}({\bf q}_i),
 \label{npsi}
 \end{equation}
With this ``wavefunction'', one can calculate the mean value of any operator 
depending on the momentum (or coordinate) of the third nucleon. We note 
that the derivative $-{\rm d}\lambda /{\rm d}E$ is always {\em positive}.

\section {Three-body forces in the dressed bag model}

 In this work, we employ the effective interaction (\ref{kernel0}) and take
into account the interaction between the dressed bag and third nucleon as an
additional three-body force (3BF). We consider here three types of 3BF:
one-meson exchange ($\pi$ and $\sigma$) between the dressed bag and third
nucleon (see Figs.3a and 3b) and the exchange by two $\sigma$-mesons, where the
third-nucleon propagator breaks the $\sigma$-loop of the two-body force
(Fig.3c).

\begin{figure}[h]
\begin{center}
\noindent {\epsfig{file=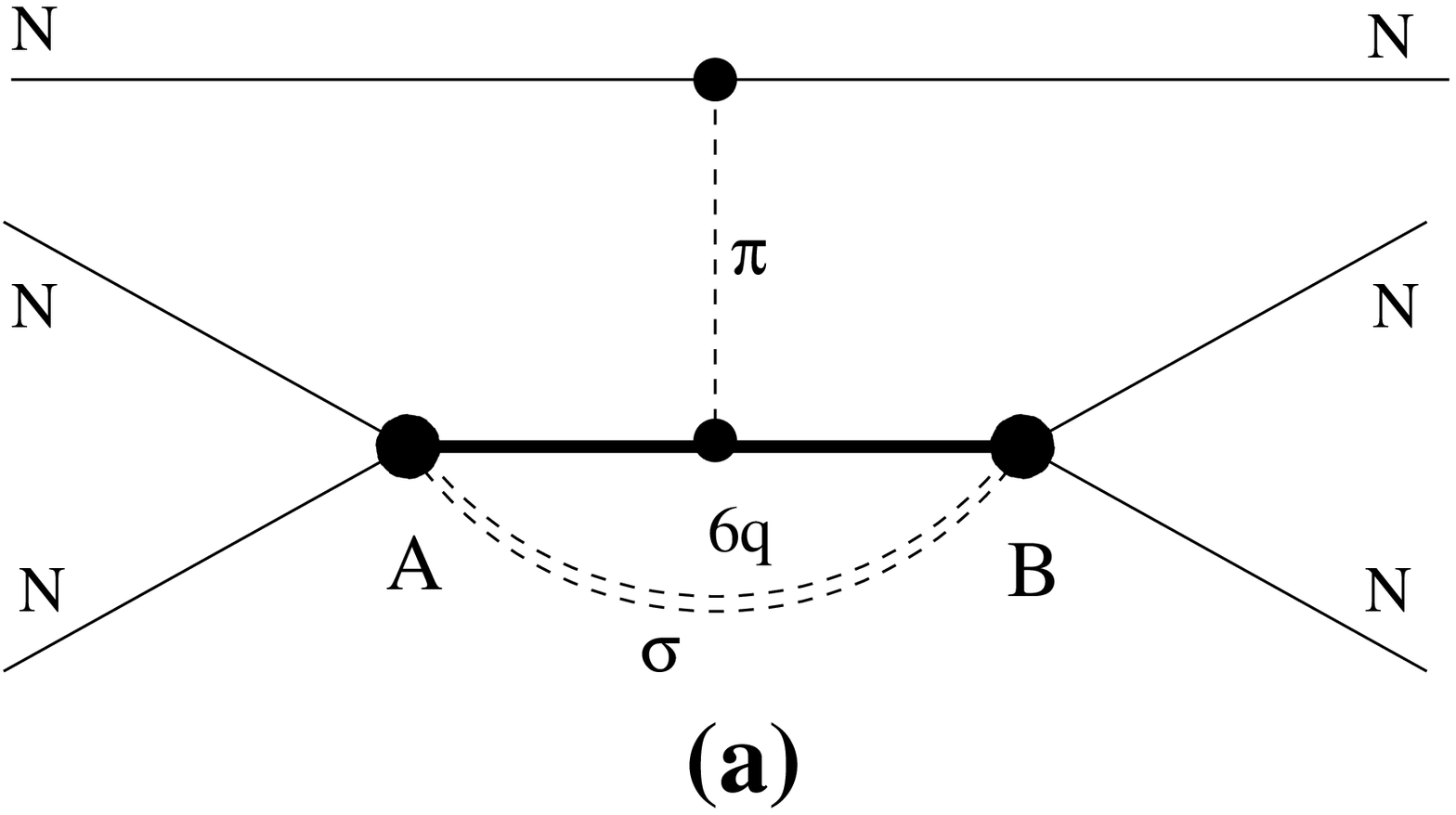,width=0.27\textwidth}}
 \hfill {\epsfig{file=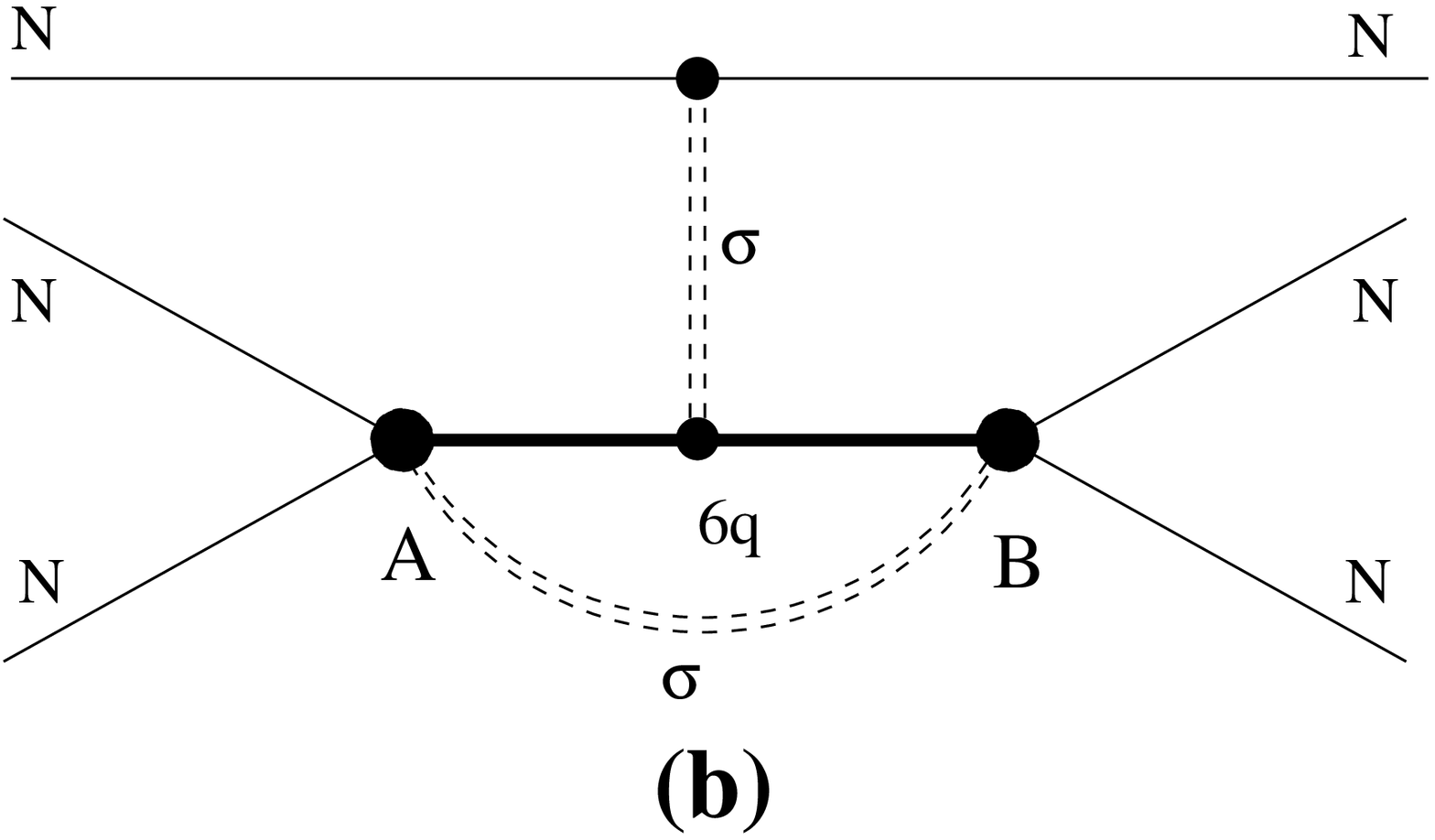,width=0.27\textwidth}}
 \hfill {\epsfig{file=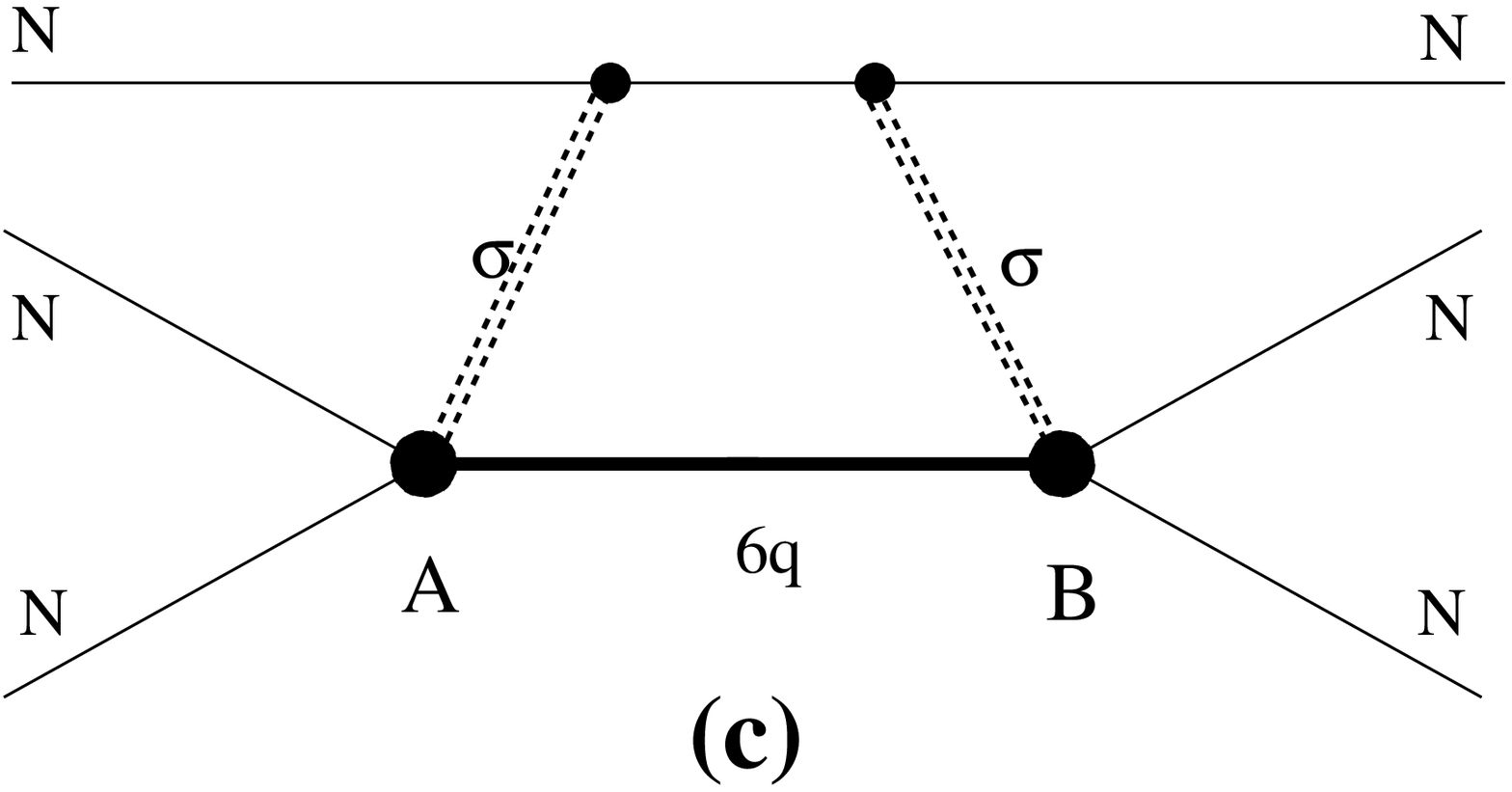,width=0.27\textwidth}}\\
\end{center}
\caption{ The graphs corresponding to three new types of three-body force}
\end{figure}

All these forces can be represented as some integral operators with
factorized kernels similar to the effective interaction
$V_{3N,6qN,3N}^{(i)} $ (\ref{kernel})
 \begin{equation}
 \fl {}^{3BF}V^{(i)}({\bf p_i, p'_i, q_i,q'_i};E)=
 \sum_{JM,J'M',L,L'}\varphi^{JM}_L({\bf p_i})\, {}^{3BF}W^{JJ'}_{LL'}({\bf q_i,q'_i};E)
 \,\varphi^{J'M'}_{L'}({\bf p'_i}),
 \label{3BF}
\end{equation}

Therefore, matrix elements for 3BF include only the overlap functions, and thus
the contribution of 3BF is proportional to the weight of the $6qN$ component in
the total wavefunction. To our knowledge, the first calculation of the 3BF
contribution  induced by one-pion exchange between the $6q$ bag and third
nucleon was done by Fasano and Lee~\cite{Fasano} in the hybrid quark-compound
bag model (QCB) using perturbation theory. They used the model where the weight
of the $6q$ component in a deuteron is ca. 1.7\%, and thus they obtained a very
small value of -0.041~MeV for the 3BF OPE contribution to the $3N$ binding
energy. Our results for the OPE 3BF agree with the results obtained by Fasano
and Lee, because the OPE contribution to 3BF is proportional to
the weight of the $6q$ component, and it should be in our case at least twice
as large as in their calculation. However, we found that a much larger
contribution comes from scalar-meson exchanges (OSE and TSE). We emphasize
that, due to (proposed) restoration of chiral symmetry in our approach, the
$\sigma$-meson mass is ca. 400~MeV, and thus the effective radius of the
$\sigma$-exchange interaction is not so small as that in conventional OBE
models. Therefore, we cannot use the perturbation theory anymore to estimate
the 3BF contribution and have to do the full calculation including 3BF in the
total three-body Hamiltonian.

\subsection{One-meson exchange between the dressed bag and third nucleon}

 For the one-meson exchange (OME) term, the three-body interaction 
 $^{3BF}W^{JJ'}_{LL'}$ takes the form:
 \[
 {}^{OME}W^{JJ'}_{LL'}({\bf q_i,q'_i};E) = \]
 \begin{equation}
  \int {\rm d}{\bf k} \frac{B^{J}_{L}({\bf k})}
  {E-m_{\alpha}-\frac{q_i^2}{2m}-\varepsilon_{\sigma}(k)}\,
  V_{OME}({\bf q_i,q'_i}) \,
   \frac{B^{J}_{L}({\bf k})}
  {E-m_{\alpha}-\frac{{q'}_i^2}{2m}-\varepsilon_{\sigma}(k)}.
\label{WOME}
\end{equation}
Therefore, the matrix element for one-meson exchange can be expressed in terms 
of the wave function of the "bag" components $\Psi^{(i)}_{6qN}$:
 \begin{equation}
 \langle \Psi_{3N} | {\rm OME}| \Psi_{3N} \rangle =
 3\langle\Psi^{i}_{6qN}
  |V^{\rm OME}|\Psi^{i}_{6qN}\rangle
 \label{OME}
 \end{equation}

 Including spin-isospin variables (it is important for evaluation of 
 pion-exchange interaction), we can 
 write down $\Psi_{6qN}^{(i)}$ as
  \begin{equation}
  \Psi_{6qN}^{(i)}=\sum_{J_iM_i,L_i}  \frac{B^{J_i}_{L_i}({\bf k}) \,
 \chi^{J_iM_i}_{L_i}(\bf q)}
  {E-E_{\alpha}-\frac{q^2}{2m}-\varepsilon_{\sigma}(k)}\,
|\alpha^{J_iM_i}\rangle\,|T_d,{\case{1}{2}}:TT_z\rangle,
  \label{PSIJL}
 \end{equation}
 where $T_d$ is the bag isospin, $T$ and $T_z$  are the total isospin of 
 the system ($T=1/2$ for $^3$H and $^3$He) and its
 $z$-projection respectively and
 $ \chi^{J_iM_i}_{L_i}(\bf q)$ are the overlap functions (\ref{ovfull}).

When calculating the matrix elements for OME, one can use a similar trick as in
the calculation of the norm for the $6qN$ component. It enables us to exclude
the vertex functions $B^{J_i}_{L_i}({\bf k}) $ from the formulas for matrix
elements. Replacing the product of propagators in the integral with respect to
the meson momentum $\bf k$ (eq.~(\ref{WOME}))  by their difference, one obtains
the following expression free of the vertex functions:

 $$
 \int \frac{B^J_L({\bf k})\,B^L_{J'}({\bf k})}
 {(E\!-\!\varepsilon(k)\!-\!\frac{q^2}{2m})
 (E\!-\!\varepsilon(k)\!-\!\frac{{q'}^2}{2m})}\,
 {\rm d}{\bf k}= \frac{\lambda^{J}_{LL'}(E\!-\!\frac{q^2}{2m})
 -\lambda^{J}_{LL'}(E\!-\!\frac{{q'}^2}{2m})}
 {q^2-{q'}^2}
 $$
 \begin{equation}
 =\Delta\lambda^J_{LL'} (q,q')
\label{repl}
 \end{equation}
This quantity is  the finite-difference analogue of the derivative 
of $\lambda$ with respect to $q^2$.

In the present calculations, we employed a rational approximation for the 
energy dependence of $\lambda^J_{LL'}$~\cite{KuInt}:
 \begin{equation}
 \lambda^J_{LL'}(E) = \lambda^J_{LL'}(0)\frac{E_0+aE}{E_0-E},
\label{approx}
 \end{equation}
where the parameters $E_0$ and $a$ are taken to be the same for all
$\lambda$'s. We found that this simple rational form can reproduce quite
accurately the exact energy dependence of the coupling constants
$\lambda^J_{LL'}(E)$ calculated from the loop diagram  in Fig.~1. For an energy
dependence such as in eq.(\ref{approx}), the energy derivative and
$\Delta\lambda (q,q')$ have the form
 \begin{equation} 
 -\frac{\rm d}{{\rm d}E}\lambda(E) = \lambda(0)\,E_0(1+a)\frac{1}{(E_0-E)^2};
\label{dlamb} 
\end{equation} 
\begin{equation} \Delta\lambda (q,q') =
\lambda(0)\,E_0(1+a)\frac{1}{E-q^2/2m}\,\frac{1}{E-{q'}^2/2m}. 
\label{ddlamb}
\end{equation} 
Thus, the matrix elements for OME can be found without
using the vertex functions $B^J_L({\bf k})$ explicitly.

\subsubsection{One-pion exchange} 
\rule{0pt}{1mm}

For one-pion exchange, we take the interaction operator in the standard form
 \begin{equation}
 V^{(i)}_{\rm OPE}= -\frac{g^2_{\pi NN}}{(2m_N)^2}({\b\sigma}^{(i)}{\bf p})
 \frac{1}{p^2+m_{\pi}^2}({\bf S_d p})({\b\tau}^{(i)}{\bf T_d}),\,{\bf p}={\bf
 q}-{\bf q'},
\label{OPE}
 \end{equation}
where ${\b\sigma}^{(i)}$ and ${\b\tau}^{(i)}$ are the spin and isospin 
variables of the third ($i$th) nucleon, whereas $\bf S_d$ and $\bf T_d$ are 
the operators of the total spin and isospin of the $6q$ bag, respectively. 
We found that the  contribution of OPE is so small that it is sufficient to 
include here only 
$S$ waves. In this case, only the central part of the OPE interaction remains:
 \begin{equation}
 V^{\rm OPE}_c= g^2_{\pi NN}\frac{m_{\pi}^2}{(4m_N)^2}
 \frac{1}{3}({\b\sigma}^{(i)}{\bf S_d})({\b\tau}^{(i)}{\bf T_d})
 \frac{1}{p^2+m_{\pi}^2}.
\label{OPECEN}
 \end{equation}
 The spin-isospin matrix element is nonzero only for a
 singlet-triplet transition:
 \begin{equation}
 \langle S_d\!=\!0,T_d\!=\!1|\frac{1}{3}({\b\sigma}^{(i)}{\bf S_d})
 ({\b\tau}^{(i)}{\bf T_d})|S_d\!=\!1,T_d\!=\!0\rangle = \frac{4}{9},
  \label{ST49}
 \end{equation}
 Then, the matrix element of the OPE contribution for $s$ waves takes the form
 \begin{eqnarray}
 \fl \langle {\rm OPE}\rangle_c =\frac{8}{3}f^2_{\pi NN}\sqrt{\lambda^0_{00}(0)
 \lambda^1_{00}(0)}\,E_0(1+a)\nonumber \\
 \times\int \frac{\chi^0_0({\bf q})}{E\!-\!E_0\!-\!\frac{q^2}{2m}}\,
 \frac{1}{({\bf q\!-\!q'})^2+m_{\pi}^2}\,
 \frac{\chi^1_0({\bf q'})}{E\!-\!E_0\!-\!\frac{{q'}^2}{2m}} \,
 {\rm d}{\bf q}\,{\rm d}{\bf q'}.
 \label{OPEfin}
 \end{eqnarray}
 Here, we take the vertex functions $B^0_0$ and $B^1_0$ differing from
 each other only by a constant. Therefore, using eq.(\ref{repl}), one can 
 exclude these functions from the formula for the matrix elements.

We did not introduce here any cut-off factor for OPE, because the overlap
 functions truncate the OPE interaction at large $q$ values. It should be noted
 that the overlap functions $\chi^0_0$ and  $\chi^1_0$ for singlet and triplet
 bags, respectively, are very similar in shape and magnitude but have {\em
 opposite signs} (see Fig.5). Therefore, the OPE  contribution (\ref{OPEfin})
 is always {\em negative}. In our case, it is equal to only $-0.1$~MeV and thus
 generally agrees with the result obtained in ref.\cite{Fasano} when the latter
 is rescaled to the larger weight of the $6q$ component in our case.

\subsubsection{One-sigma exchange (OSE)}
\rule{0pt}{1mm}

The scalar meson exchange operator does not include any spin-isospin
variables. Therefore, eq.(\ref{WOME}) for $W^{JJ'}_{LL'}$ can be
simplified and, in view of the energy dependence given in
eq.~(\ref{approx}), reduces to the form
 \[
 {}^{OSE}W^{JJ'}_{LL'}({\bf q_i,q'_i};E) = \delta_{JJ'}
 \lambda^J_{LL'}(0)\,E_0(1+a) \]
 \begin{equation}
  \times \frac{1}{E\!-\!E_0\!-\!\frac{q_i^2}{2m}}\,
  \frac{-g^2_{\sigma NN}}{({\bf q}_i-{\bf q}'_i)^2+m_{\sigma}^2}\,
  \frac{1}{E\!-\!E_0\!-\!\frac{{q'}^2_i}{2m}}
\label{WOSE}
\end{equation}
Thus, the matrix element of OSE 3BF is determined by the formula
 \begin{eqnarray}
 \langle {\rm OSE}\rangle =-3g^2_{\sigma NN}\sum_{JM,L,L'}
 \lambda^{J}_{LL'}(0)\,E_0(1+a)\,\nonumber\\
   \times \int \frac{\chi^{JM}_{L}({\bf q})}{E\!-\!E_0\!-\!\frac{q^2}{2m}}\,
 \frac{1}{({\bf q\!-\!q'})^2+m_{\sigma}^2}\,
 \frac{\chi^{JM}_{L'}({\bf q'})}{E\!-\!E_0\!-\!\frac{{q'}^2}{2m}}\,
 {\rm d}{\bf q}\,{\rm d}{\bf q'}.
 \label{OSE}
 \end{eqnarray}
 In the actual calculations, we used the ``light'' $\sigma$-meson mass
 $m_{\sigma}= 400$~MeV.

 The physical meaning of the OSE three-body force is easily understood: it
 represents the interaction of the third nucleon with the $\sigma$-meson cloud
 of the bag. In view of the enhancement of the $\sigma$-field around the dense
 $6q$ bag, this contribution is very important for our understanding of the
 properties of $3N$ and heavier nuclei.

\subsection{Two-sigma exchange (TSE)}

The two-sigma process shown in Fig.4 also contributes significantly to 3BF.
This $3N$ interaction seems less important than the OSE force, because this
interaction imposes a specific kinematic restriction on the $3N$
configuration.\footnote{It follows from the intuitive picture of this
interaction that this force can be large only if the momentum of the third
nucleon is almost opposite to the momentum of the emitted $\sigma$-meson. Thus,
a specific $3N$ kinematic configuration is required when two nucleons approach
close to each other to form a bag, while the third nucleon has a specific space
localization and momentum.}

\begin{figure}
\centerline{\epsfig{file=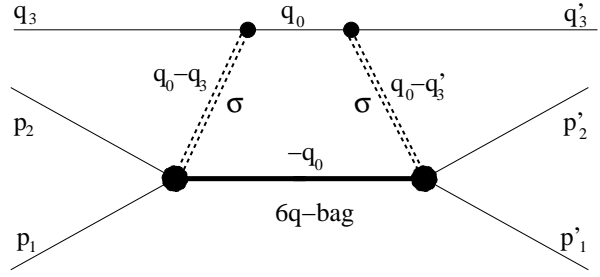,width=0.5\textwidth}}
\caption{Three-body force due to two-sigma exchange}
\end{figure}

The operator of this interaction includes the vertex functions for the ($NN \to
6q\!+\!\sigma$) and reverse transitions so that these vertices cannot be
excluded similarly to the case of OME. Therefore, we have to choose some form
of these functions. It is naturally to suppose that these vertices are the same
as those assumed in two-body DBM; i.e., they can be normalized by means of the
coupling constants $\lambda(E)$, which, in turn, are chosen  in the two-nucleon
sector to accurately describe $NN$ phase shifts and deuteron properties (see
eq.(\ref{adjust}) for vertex normalization). We use the Gaussian form factor
for these vertices:

\begin{equation}
B^J_L({\bf k})=B_0^{JL}\frac{{\rm e}^{-b^2k^2}}{\sqrt{2\omega_{\sigma}(k)}},
 \label{Bk}
 \end{equation}
where $\bf k$ is the meson momentum and the parameter $b$ is taken from 
the microscopical quark model~\cite{KuInt}:
\begin{equation}
b^2=\frac{5}{24}b_0^2;\qquad b_0=0.5\mbox{ fm}.
 \label{b0}
 \end{equation}
 Then, the vertex constants $B_0$ should  be found from the equation:
 \begin{equation}
 \frac{1}{(2\pi)^3}\int {\rm d}{\bf k}\frac{B_0^{JL}B_0^{JL'}{\rm e}^{-2b^2k^2}}
 {(E\!-\!m_{\alpha}\!-\!\varepsilon_{\sigma}(k))\,2\omega_{\sigma}(k)}
 = \lambda^{J}_{LL'}(E),
 \label{adjust}
 \end{equation}
where $\lambda^{J}_{LL'}(E)$ are the coupling constants employed in the
construction of the  DBM in the  $NN$ sector and are fixed by $NN$ phase
shifts.

For the $\sigma NN$ vertices, we also take the Gaussian form factor:
\begin{equation}
g_{\sigma NN}\,{\rm e}^{-\alpha^2k^2}, \mbox{ with } \alpha^2=\frac{1}{6}b_0^2.
 \label{sigmaNN}
 \end{equation}
Then, the box diagram in Fig.~4 can be expressed in terms of the integral with 
respect to the momentum $\bf q_0$ of the third nucleon in the intermediate 
state: 
\[
 {}^{TSE}W^{JJ'}_{LL'}({\bf q,q'};E) = \delta_{JJ'}
  g_{\sigma NN}^2\,B_0^{JL}\,B_0^{JL'}\,
\]
 \begin{equation}
  \times\frac{1}{(2\pi)^3}\int {\rm d}{\bf q_0}
  \frac{{\rm e}^{-(\alpha^2\!+\!b^2)({\bf q_0\!-\!q})^2}}
  {m_{\sigma}^2+({\bf q_0\!-\!q})^2}\,
  \frac{1}{E\!-\!m_{\alpha}\!-\!\frac{q_0^2}{2m}}\,
  \frac{{\rm e}^{-(\alpha^2\!+\!b^2)({\bf q_0\!-\!q'})^2}}
  {m_{\sigma}^2+({\bf q_0\!-\!q'})^2}.
\label{WTSE}
\end{equation}
Thus, the matrix element for the total contribution of TSE takes the form
\begin{equation}
 \langle {\rm TSE}\rangle =3\,\sum_{JM,L,L'}
 \int \chi^{JM}_{L}({\bf q})\, {}^{TSE}W^{JJ}_{LL'}({\bf q,q'};E)
 \chi^{JM}_{L'}({\bf q'})\,
 {\rm d}{\bf q}\,{\rm d}{\bf q'}.
 \label{TSE}
 \end{equation}
After the partial wave decomposition, these six-dimensional integrals can be 
reduce to
two-dimensional integrals, which are computed numerically by means of the
Gaussian quadratures.

\section{Results} 
The present calculations of $3N$ bound states have been made with the two
variants of the DBM, versions I and II, one of which being used in our previous
two-nucleon studies~\cite{KuInt}. For the other variant, we use a weaker energy
dependence $\lambda(E)$, which is taken directly from the loop integral
(eq.~(\ref{adjust})) corresponding to the diagram in Fig.~1. The effective
three-nucleon Hamiltonian in the $3N$ channel includes both two-body and
three-body forces (although, as was noted above, such a division is justified
only for the Yukawa ($t$-channel) exchange contribution):
\begin{equation}
  {\cal H}_{3N}=T+\sum_{i=1}^3 V^{NN}_i + \sum_{i=1}^3 V^{3BF}_i.
 \label{H3NM}
 \end{equation}
 
The parameters of two-body potentials $V^{NN}_i $ (for the two versions)
provide a quite good description of the deuteron observables $NN$ and phase
shifts in the wide energy interval 0 -- 1~GeV.
The parameters of the three-body force $V^{3BF}_i$ were discussed in the
previous section. We took into account both the OSE and TSE contributions to
3BF. The contribution of OPE interaction between the dressed bag and third
nucleon to the $3N$ binding energy was found to be only about -0.1~MeV.
Therefore, we did not include this term in the complete variational calculation
but estimated it with the perturbation theory (PT). The present $3N$
calculations were made using a highly accurate variational method and the
antisymmetrized Gaussian basis~\cite{Tur1,Tur2} with non-linear scale
parameters. This basis is extremely flexible and has an enormous advantage for
such tedious  calculations.  It enables us to obtain completely analytical
formulas for the matrix elements of effective two- and three-body forces,
except two-sigma exchange (these formulas will be presented in another paper).
In the present calculations, we used the variational basis including even
partial waves with the total orbital angular momenta $L=0,2$ and took into
account the $NN$ force components for the $^1S_0$ and $^3S_1-{}^3D_1$ channels.
The dimension of the basis was increased until the converged results were
reached. The results presented in Table~2 are obtained with the basis that
includes six three-body partial components and has the total dimension $N=688$.

\begin{table}
\caption{ Results of $3N$ calculations with two- and three-body forces
for two versions (I and II) of the dressed-bag model.}

\medskip

\begin{tabular}{|*{9}{c|}}  \hline
  & $E$& $P_D$ \% &  $P_{6qN}$ \%
  & \multicolumn{5}{|c|}{Individual contributions to $H$, MeV} \\ \cline{5-9}
 & MeV&&&$T$& $V^{NN}_{NqN}$&
 $V^{3BF}_{OSE}$& $V^{3BF}_{TSE}$& $V^{3BF}_{OPE}(PT)$  \\ \hline
\multicolumn{9}{|c|}{ Version I, $g_{\sigma NN}=9.5$}\\ \hline
   $^3$H$^{(1)}$ & -5.83 & 6.44 &  7.87 & 75.0 &-64.45
  &-1.54 (PT)$^*$& -0.43 (PT)$^*$ & \\ \hline
 $^3$H$^{(2)}$ & -4.14 & 5.81 &  6.24 & 58.1 &-49.88 &-0.83 (PT)$^*$& -0.22 (PT)$^*$ & -0.01 \\ \hline
 $^3$H$^{(3)}$  & -8.326 &6.84  & 10.84 &110.7 &-92.9 &-4.72 & -2.03 & -0.1  \\ \hline
 $^3$He$^{(3)}$ & -7.588 & 6.80  & 10.66 &108.2 &-91.0 &-4.55 & -1.99 & -0.1  \\ \hline \hline
\multicolumn{9}{|c|}{ Version II, $g_{\sigma NN}=8.6$}\\ \hline
   $^3$H$^{(1)}$ & -6.12 & 6.67 &  5.45 & 79.0 &-67.82
  &-1.30 (PT)$^*$& -0.56 (PT)$^*$ & \\ \hline
 $^3$H$^{(2)}$&-5.01 & 6.23 &  4.67 & 66.9 &-57.3 &-0.87 (PT)$^*$& -0.36 (PT)$^*$ & -0.01 \\ \hline
 $^3$H$^{(3)}$&  -8.358 & 7.06 & 7.31 & 110.8 & -94.2 & -2.79 & -1.66  & -0.1  \\ \hline
 $^3$He$^{(3)}$ &  -7.565 & 7.00 & 7.14 & 107.8 & -91.8 & -2.66 & -1.58  & -0.1  \\ \hline
\end{tabular}\\[0.5 mm]
{\small
 $^{(1)}$ -- without 3BF, without $q^2$-dependence,\\
 $^{(2)}$ -- without 3BF but with $q^2$-dependence, \\
 $^{(3)}$ -- with 3BF and with $q^2$-dependence,\\
 $^*$ These values have been found with the $3N$ wavefunctions in those
calculation the 3BF have been omitted, i.e. within PT. The comparison of these
entries with those found in the complete calculation (see rows denoted $^3$H$^{(3)}$) 
shows with evidence inapplicability of the PT. 
}
\end{table}

Table~2 presents our results for the following static properties of $^3$H:
binding energy $E$, $D$-wave percentage $P_D$, 
and weight of the dressed "bag" $6qN$ component $P_{6qN}$. In addition, this
table shows the contributions of the following individual parts of the
Hamiltonian to the total three-body expectation value: three-nucleon kinetic
energy $T$, two-body effective force $V^{NN}_{NqN}$, and three-body force due
to one-sigma ($V^{3BF}_{OSE}$) and two-sigma exchanges ($V^{3BF}_{TSE}$). The
contribution of 3BF due to OPE calculated with perturbation theory (PT) is also
given in Table~2. The other details of the results presented in the Table,
especially related to the $^3$He properties, are discussed in the forthcoming 
paper.

For both variants of the model, we present also the result calculated
disregarding both 3BF and the $q^2$ dependence of the effective two-body force
$V_{NqN}$ on the momentum of the third nucleon (see the first and fifth rows of
Table 2). The results in the second and sixth rows of Table 2 are obtained
including the $q^2$ dependence of $V_{NqN}$, but disregarding 3BF. The
percentages of the $D$-wave $P_D$ and $6qN$ component $P_{6qN}$ given  in
Table~2 were obtained with incorporation of the three $6qN$ components; i.e.,
these values correspond to the normalization of the total (four-component)
wavefunction of the system to unity.

\section{Discussion}
The $3N$ results presented in the previous section differ significantly both
from the $3N$ results found with the conventional models for $NN$ and $3N$
forces (based on Yukawa's meson exchange mechanism) and from the results 
obtained in the framework of hybrid models~\cite{hyb}, which include the
two-component representation of the $NN$ wavefunction
$\Psi=\Psi_{NN}+\Psi_{6q}$. It is convenient to discuss these differences in
the following order.
 \begin{itemize}
 \parindent=2em
 \item[(i)] 
 We found that the $q^2$ dependence of pair $NN$ forces on the momentum of the
 third particle in the $3N$ system is more pronounced in our case than in other
 hybrid models~\cite{hyb,Bakk,Weber,Sim}: the $3N$ binding energy decreases by
 more than 1.7~MeV, from 5.85 to 4.14 MeV (cf. the first and second rows in
 Table~2). From the more general point of view, it means that, in our approach,
 pair $NN$ interactions (except Yukawa OPE and TPE terms), being ``embedded''
 into a many-body system, loose their two-particle character and become
 substantially many-body forces (depending on the momenta of other particles of
 the system).

\item [(ii)] 
Due to such a strong $q^2$ dependence (of ``repulsive'' character), the $3N$
system calculated including only pair forces turns out to be strongly
underbound ($E=-4.14$~MeV). In other words, the``pairwise'' $NN$ forces
(including their $q^2$ dependence on the momenta of the third nucleon) give
only about half the total $3N$ binding energy, leaving the second half for the
3BF contribution. Therefore, the following question is decisively important:
can the three-body force (inevitably arising in our approach) give the large
missing contribution to the $3N$ binding energy? Usefulness of the developed
model for the description of nuclear systems depends directly on the answer to
this important question. It is appropriate here to remind that in the
conventional 3BF models such as Urbana-Illinois or Tucson-Melbourne, the
contribution of 3BF to the total $3N$ binding energy does not exceed 1~MeV;
i.e., this contribution can be considered as some correction ($\sim 15$\%),
although it is significant for the precise description of the $3N$ system.

 \item[(iii)] 
 Fortunately, the contribution of 3BF induced by $\sigma$ and $2\sigma$
exchanges enables one to fill this 4.4~MeV gap between the two-body force
contribution and experimental value. In fact, including both one- and two-sigma
exchange contributions to 3BF, taken with {\em the same parameters as in the
initial $NN$-force model} and a quite reasonable coupling constant $g_{\sigma
NN}=8 \div 9.5$, we obtain the $3N$ binding energy that is very close to the
experimental value (see rows 3,4 and 7,8 in Table~2). Thus, the presented force
model leads to a very reasonable binding energy for the three-nucleon system
but with the strongly enhanced (as compared to the traditional $3N$ force
model) contribution of three-body forces.

\item[(iv)] 
The contributions of pair and different three-body forces to the total $3N$
binding energy for $^3$H are given in the third and fourth rows of Table~2.
From the results presented in these rows, one can conclude that the total 3BF
contribution to the $3N$ binding energy dominates and, in fact, determines the
structure of the $^3$H ground state\footnote{It should be noted that the
relative contribution of the pair effective force $V_{NqN}$ to the binding
energy decreases noticeably when including 3BF (due to strengthening the $q^2$
dependence of pair forces).}. Moreover, comparing the third and fourth rows of
the table, one can see a ``nonlinear'' effect of self-strengthening  for the
3BF contribution. The comparison of the results presented in the third and
fourth rows of the table (see the second, fourth, and fifth columns) shows
clearly that the binding energy is almost proportional to the weight $P_{6qN}$
of the $6qN$ component in the total $3N$ wavefunction. Thus, when the  weight
of the $6qN$ component increases, the 3BF contribution, which is related
directly only to this component of the total wavefunction, increases 
accordingly. However,
the enhancement of the pure attractive 3BF contribution squeezes the $3N$
system and thus reduces its rms radius, i.e., the mean distance between
nucleons, which, in turn, again increases the weight of the $6qN$ component. In
other words, a some chain process strengthening the attraction in the system
arises. This process is balanced both by the weakening of the effective pair
interaction due to the $q^2$ dependence and by the repulsive effect of the
orthogonalizing pseudopotentials included in each pair interaction.

There are two another important stabilizing factors weakening the strong
three-body attraction in the $3N$ system. First, the generation of the
short-range repulsive vector $\omega$-field, where all three nucleons are close
to each other~\cite{Bled}. Since the $\omega$-meson is heavy, this field is
located in the deep overlap region of all three nucleons. In this study, we
omitted the three-body contribution of this repulsive $\omega$-field. This
repulsive contribution will keep the whole system from the further collapse due
to the strong attractive $3N$ force induced by the scalar field.

The second factor slightly weakening the effective $3N$ attraction is
associated with the conservation of the number of scalar mesons generated in
the $2N$ and $3N$ interaction process. The problem is that $2\sigma$-exchange
giving the 3BF contribution (see Fig.~4) arises due to the break of the
$\sigma$-meson loop, which induces the main $2N$ force. In other words, the
$\sigma$-meson generated in the transition of pair nucleons from the $NN$ phase
state to the $6q$ state is absorbed either in the $6q$ bag with closing the
loop or by the third nucleon resulting in the 3BF contribution. Thus, the
appearance of such a 3BF should weaken attraction between nucleons in the pair.
We carefully estimated the effect of the meson-number conservation for the TSE
contribution on the total $3N$ binding energy. Its magnitude occurred to be
rather moderate in the absolute energy scale (ca. 0.3 -- 0.4~MeV) but quite
noticeable within the whole TSE contribution. However, when the total nucleon
density increases (and the relative TSE contribution also increases), the
effect is enhanced.

\item[(v)]
Dependence of the two-body coupling constants $\lambda (\varepsilon)$ upon the
average momentum of other nucleon in three-nucleon system (see e.g.
eq.~(\ref{kernel0}) can be interpreted generally as a density dependence of the
two-body force in many-nucleon system. Thus, one fixes (e.g. at energy
$\varepsilon = \varepsilon_d$) the value of energy dependent coupling constant
$\lambda(E-q^2/2m)$ of our two-body force, i.e., if one disregards its
$q^2$-dependence (this $q^2$-dependence leads to a weakening of the two body 
force in a many-nuclear system when 
$q^2$ is rising), than the neglected $q^2$-dependence must be compensated by an
additional {\em repulsive
density-dependent effective three-body force}.

On the other hand, it is well known from the Skirme-model calculations of
nuclei that just similar repulsive phenomenological density-dependent
three-body force should be added to conventional $2N$- and $3N$-forces to
guarantee the saturation properties of heavy nuclei. Thus, in this respect the
present force model is also in a qualitative agreement with phenomenological
picture of nuclear interactions.

\begin{figure}
 \centerline{ \epsfig{file=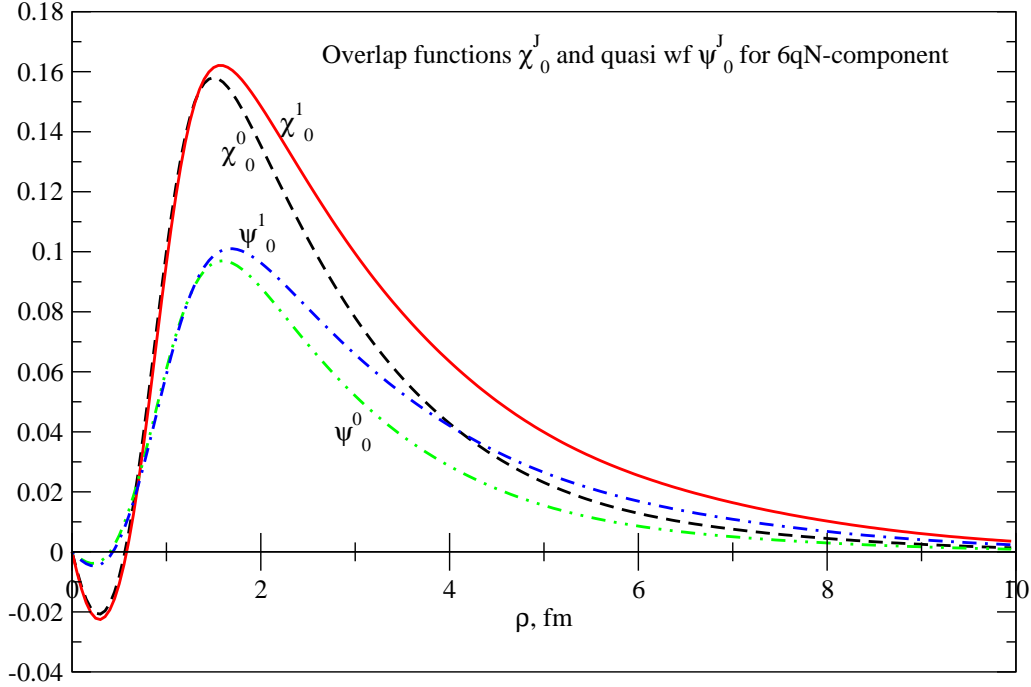,width=0.8\textwidth,angle=-90}}
 \caption {$6qN$ component of the $^3$H wavefunction:  the 
triplet and singlet $s$-wave overlap functions $\chi^1_0(\rho)$ (solid line) 
and $\chi^0_0(\rho)$ (dashed line) and the corresponding
``quasi-wavefunctions'' $\tilde{\psi}^1_0(\rho)$ (dash-dotted line) and 
$\tilde{\psi}^0_0(\rho)$ (dot-dot-dashed line). 
The triplet functions are multiplied by a factor of (-1).}
\end{figure}

\item[(vi)] Figure 5 shows two types of the $^3$H $s$-wave wavefunctions in 
the $6qN$ channel: the triplet $\chi^1_0(\rho)$ (solid line) and singlet
$\chi^0_0(\rho)$  (dashed line) overlap functions  given by
eqs.(\ref{ovJL}, \ref{ovfull}) and the corresponding ``quasi-wavefunctions'' 
$\tilde{\psi}^0_0(\rho)$ and $\tilde{\psi}^1_0(\rho)$  normalized according to
eq.~(\ref{npsi}).  The triplet functions are multiplied by (-1) for
convenience.
It is evident that there is an inner node (at $r_n\simeq 0.56$~fm) in all
$s$-wave components of the quasi-wavefunctions and the overlap functions. These
inner nodes are stationary when the total energy increases and the $Nd$
scattering problem is considered. As is seen in Fig.~5, the overlap function
$\chi^1_0(\rho)$ for the triplet $6qN$ channel has a more extended tail than
the singlet component $\chi^0_0(\rho)$. This is due to the fact that the $3N$
system in the spin-triplet channel (coinciding with deuteron channel) is less
bound, by about 2~MeV, than in the spin-singlet channel.

\end{itemize}

\section{Conclusion}
In this paper, we have developed a formalism for a multi-component description
of the three-nucleon system within the new approach to the $2N$ and $3N$
interactions based on the dressed dibaryon and $\sigma$-field generation. It
has been shown that the DBM applied to the $3N$ system results automatically in
a new three-body scalar force due to the interaction between the dressed
dibaryon and third nucleon. This force plays a crucial role in the structure of
few-nucleon systems. Our accurate variational calculations have demonstrated
that new 3BF gives  half the 3N binding energy, whereas the 3BF contribution in
the traditional $NN$-force approaches gives about 15\% of the total binding
energy. Thus, the suggested approach to the $NN$ interaction can lead to
significant revision of relative contributions of two- and many-body forces in
nuclear systems.

It is crucially important that the DBM gives an 8 - 11\% non-nucleonic
component in the 3N wavefunction, while this component in the deuteron is equal
to only about 3\%, leading to a reformulation of many effects in few-nucleon
systems and other nuclei as well. It is probable that the weight of such 
non-nucleonic components in heavy  nuclei can be even higher with an increase
in the mass number and nuclear density. 

 Generalization to $4N$ (and more) systems can be done rather similarly  to the
case of conventional force model. Having $2N$- and $3N$-forces fixed  from
$2N$- and $3N$- calculations one can straightforwardly consider  systems with
larger number of nucleons with these forces to be incorporated,  i.e. ignoring
the  contribution of $4N$ etc. forces. However there is still another
possibility  to proceed for many-body systems. If the contribution of such
many-body forces will occur to be significant one can reformulate the whole
scheme in terms  of collective $\sigma$- (and other) fields interacting with
relativistic  nucleons and dibaryons, i.e. in the spirit of Walecka-Serot model
of  hadrodynamics. The only essential difference in this point from the 
Serot-Walecka model is adding dibaryon components to the nucleonic ones. 
Moreover, our approach, contrary to the conventional OBE-approach, leads
inevitably to a strong enhancement of collective  $\sigma$-field in nuclei and
thus to validity of Walecka-type model.

 Numerous modern experiments could corroborate these results. In particular,
according to the recent experiments $^3{\rm He}(e,e'pp)$\cite{Nikef} and their
theoretical interpretation on the basis of fully realistic $3N$ calculations,
the cross sections for the $^3{\rm He}(e,e'pp)$ process cannot be explained
within a fully realistic $3N$ model incorporating the process where a proton
knocked out by a virtual $\gamma$-ray photon is sequentially rescattered by the
second nucleon. These calculations show that the leading contribution comes
from the one-step process where a virtual photon is absorbed by one nucleon in
the target, while the remaining two nucleons (which are spectators in this
process) are emitted (in coincidence) at the second step. This important
conclusion has been further corroborated in recent experiments at the Jefferson
Laboratory when the incident electron beam energy has been increased up to
$E_e=2.2$~GeV and 4.4~GeV~\cite{Jef}. The data of the two different experiments
give a clear evidence of very strong short-range $NN$ correlation in the $^3$He
ground state. This correlation still cannot be explained within the traditional
pattern for the $3N$ system. In addition, our approach has recently been
partially corroborated~\cite{Kask} from the other side by considering a model
for $2\pi$ production in $pp$ collisions at $E_p=750$ and 900~MeV. The authors
have found that almost all particle energy- and angular correlations (e.g.
$\pi^+ \pi^-,\, pp, \, \pi pp$ and etc.) can be explained quantitatively by
assuming that $\pi^+ \pi^-$ production occurs through the generation of an
intermediate light $\sigma$-meson with the mass $m_{\sigma}\simeq 380$~MeV and
rather narrow width. These values generally
agree with the parameters adopted in our $NN$ model~\cite{KuJPG,KuInt} and
drastically disagree with the values assumed in OBE and other potential models.

Very interesting general implication of the results presented here is their
evident interrelation to the famous Walecka hadrodynamic model for
nuclei~\cite{Walecka}. It is well known that the Walecka model describes nuclei
and nuclear matter in terms of the scalar $\sigma$ and vector $\omega$-fields,
where the $\sigma$-field gives the attractive contribution, while the vector
$\omega$-field balances this attraction by short-range repulsion. It is very
important that both basic fields exist (in the model) as the {\em explicit
degrees of freedom} (together with relativistic nucleons) in contrast to
conventional meson-exchange models for nuclear forces, where mesons appear as
the carriers of forces rather than as the {\em explicit} field degrees of
freedom. Our approach does include the $\sigma$-meson (and potentially the
$\omega$-meson) degrees of freedom in an explicit form similarly to the Walecka
model. Moreover, since the average kinetic energy of the $3N$ system is high in
our model (it is higher than that in the conventional OBE approach by a factor
of more than 2), nucleon motion is closer to the relativistic case, and thus
the similarity with the Walecka model is even closer.

These general arguments give an additional strong support for the $2N$- and
$3N$-force model presented here. Quite independent numerous arguments in favour
of this approach in nuclear physics are presented in the subsequent
paper~\cite{Coul}.

\ack
We are deeply grateful to many our colleagues for continuous
encouraging in the course of this work and to the staff of Institut f\"ur
Theoretische Physik der Universit\"at T\"ubingen where the most part of
calculations has been performed.  This work was supported in part by
der Deutsche Forschungsgemeinschaft (grant No. Fa-67/20-1) and the
Russian Foundation for Basic Research (grants No.01-02-04015 and 02-02-16612).

\end{document}